\documentclass{aa} 
\usepackage{graphicx} 
\newcommand{\lae}{\mathrel{<\kern-1.0em\lower0.9ex\hbox{$\sim$}}}
\newcommand{\gae}{\mathrel{>\kern-1.0em\lower0.9ex\hbox{$\sim$}}}

   \title{{\it INTEGRAL} observations of the Be/X-ray binary EXO 2030+375 during outburst} 
 
   \subtitle{} 
 
   \author{A. Camero Arranz
         \inst{1} 
          \and 
          C.A. Wilson
          \inst{2} 
	    \and 
    	    P. Connell
          \inst{1} 
          \and
	    S. Mart\'{\i}nez N\'u\~nez 
          \inst{1} 
	    \and  
	    P. Blay
          \inst{1} 
	    \and	          
	    V. Beckmann
          \inst{3,4}
            \and\linebreak
	    V. Reglero
          \inst{1} 
          } 
 
\institute{ $^1$   GACE, Instituto de Ciencias de los Materiales, Universidad de Valencia,P.O. Box 20085, 46071 Valencia, Spain \\
$^2$  XD12 NASA Marshall Space Flight Center, Huntsville, AL 35812, USA \\     
$^3$  NASA Goddard Space Flight Center, Code 661, Greenbelt, MD 20771, USA\\	 
$^4$  Joint Center for Astrophysics, Depart. of Physics, University of Maryland, Baltimore County, MD 21250, USA}

\authorrunning{Camero Arranz et~al} 
\titlerunning{{\it INTEGRAL} observations of the Be/X-ray binary EXO 2030+375 during outburst} 
 
\offprints{A.Camero Arranz\\ \email{Ascension.Camero@uv.es}} 
 
\date{Received  29 April 2005/ Accepted 9 June 2005} 

\begin{document}

\abstract{ 
We present a type--I outburst of the high-mass X--ray binary EXO 2030+375,  
detected during {\it INTEGRAL}'s Performance and Verification phase in December 2002 
(on--source time about $10^{6}$ seconds). In addition, six more outbursts have been observed 
during {\it INTEGRAL}'s Galactic Plane Scans. X-ray pulsations have been detected with a  
pulse period of 41.691798$\pm$0.000016 s. The X-ray luminosity in the 5--300\rm\,keV energy range  was  
9.7$\times10^{36}$ erg s$^{-1}$, for a distance of 7.1 kpc. Two unusual features were found in the light 
curve, with an initial peak before the main outburst and another possible spike after the 
maximum. {\it RXTE} observations confirm only the existence of the initial spike. Although the initial peak appears 
to be a recurrent feature, the physical mechanisms producing it and the possible second spike are unknown. Moreover,  a  
four-day delay between periastron passage and the peak of the outburst is 
observed. We present for the first time a  5--300 keV broad-band spectrum of this source. It can 
be modelled by the sum of a disk black body (kT$_{BB} \sim$8 keV)  with either a power law model with 
$\Gamma$=2.04$\pm$0.11 keV or a Comptonized component  (spherical geometry, kT$_e$=30 keV, $\tau$=2.64, kT$_W$=1.5 keV).

}
 
\maketitle
 
\section{Introduction} 
 
The high-mass X-Ray binary EXO 2030+375 is an X-ray \linebreak transient system that 
consists of  a neutron star (NS) \linebreak orbiting a Be companion, hence forming a 
Be/X-ray \linebreak binary. Be stars are  rapidly rotating objects with a 
quasi-Keplerian disk around their equator. The ultimate cause of the 
formation of the disk is still under heavy debate, but the high 
rotation velocities of these types of stars must play an important role  
(Townsend 2004). The optical and infrared emission is dominated by the 
donor  star and  characterised by spectral lines in emission (particularly 
those of  the Balmer series) and IR excess. The standard model of a 
Be/X-ray binary ascribes the high-energy  radiation  \linebreak to an accreting 
mechanism that takes place when the compact object interacts with the  Be 
star's circumstellar disk, giving rise to an X-ray outburst. 
 
EXO 2030+375 was discovered by  {\it EXOSAT} in 1985 during a giant (or type II) 
X-ray outburst (Parmar et al. 1989), probably arising from a dramatic 
expansion of the disk  surrounding the Be star (Coe 2000). The X-ray luminosity in 
the energy range 1--20 keV reached a value of $L_x\sim10^{38}$erg s$^{-1}$, and  
a large intrinsic spin-up  rate (Reynolds et al. 1993), as 
well as 0.2 Hz QPO (Angelini, Stella, \& Parmar 1989) were observed. Since 
then, only normal (or type I) outbursts have been detected:  {\it EXOSAT} (28 October - 3 November 
 1985, Parmar et al. 1989), {\it Ginga} (29-31 October 1989 and 24 October 1991; Sun et al. 1992), {\it ROSAT} 
(November 1990; Mavromatakis 1994), {\it CGRO}/BATSE and {\it RXTE} satellites  (since April 1991 
outbursts were detected at most periastron passages and continue to be detected with the ASM, Wilson 
et al. 2002, 2005), and {\it INTEGRAL} (6-18 December 2002, and continue to be detected during GPS observations).  These 
type I \linebreak outbursts have $L_x <10^{37}$ erg s$^{-1}$, are modulated with the orbital period and display low spin-up 
rate episodes.

With an orbital period of $P_{\rm orb}$=46.0214$\pm$0.0005 d   and an eccentricity  of $e$=0.419$\pm$0.002  (Wilson et al. 2002),  EXO 2030+375 exhibits variability  on all  time scales. Its optical counterpart  is a Be ($V=20\rm\,mag$) main-sequence star  (Motch \& Janot-Pacheco 1987; Coe et al. 1988). The optical/IR emission seems to be related to the activity of the Be star's disk (Reig et al. 1998). In the X-ray band the neutron star shows 41.7-s pulsations  (Parmar et al. 1989; Reynolds, Parmar \& White 1993, Reig \& Coe 1998;  Wilson et al. 2002, 2005), and active/inactive periods are mainly related to the periastron  passage (46.02 d).  The spectral shape of EXO 2030+375 in the 2--20 keV energy range has been modelled by an exponential cut-off ($\Gamma=1.00\pm 0.06$) plus absorption  and an iron emission line at  $\sim$ 6.5 keV  (Reig \& Coe 1999). In the 20--150 keV energy band Stollberg et al. (1999) found  a thermal bremsstrahlung model to be the best fit ($kT=20.2\pm 0.3\rm\,keV$).

In this paper we present for the first time a broad band spectrum of  
EXO 2030+375 using all three high-energy instruments on-board {\it INTEGRAL}.
In addition, we have analysed {\it RXTE} data from three other outbursts. Furthermore, a detailed  timing analysis 
of this source is shown. {\it INTEGRAL} preliminary results of EXO 2030+375 have been reported by
Mart\'{\i}nez N\'u\~nez et al. (2003), Kuznetsov et al. (2003, 2004), Bouchet et al. (2003) and Camero et al. (2004).

%_____________________________________________________________
%                        figure with caption on the right side
%-------------------------------------------------------------

 \begin{figure}
  \centering
  \includegraphics[height=5.5cm,width=7.9cm]{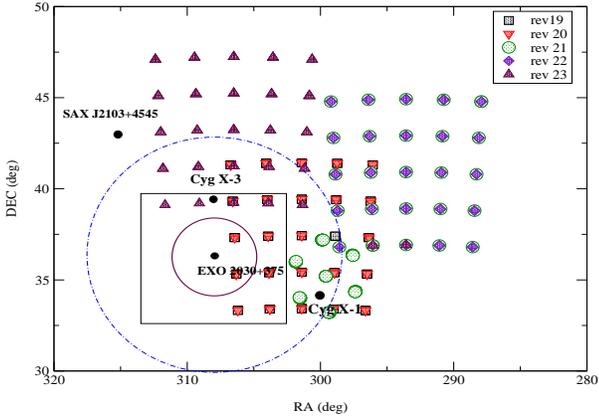}
    \caption{Dithering pattern (5x5) during PV phase around Cygnus region. From inside to outside, the fully coded fields of view of JEM-X, ISGRI and SPI instrumemts, respectively, are also shown.\label{imag}}
 \end{figure}

\section{Observations and data reduction}

The \textbf{INTE}rnational \textbf{G}amma-\textbf{R}ay \textbf{A}strophysics  
\textbf{L}aboratory ({\it INTEGRAL}, Winkler et al. 2003) consists of three  
coded mask telescopes: the spectrometer SPI (20 keV--8 MeV), the imager IBIS 
(15 keV--10 MeV), and the X-ray monitor JEM-X (4--35\rm\,keV), as well as the 
optical monitoring camera OMC (V,  500--600nm). 
 
The spectrometer SPI has an angular resolution of $2.8^\circ$ and an excellent 
energy resolution (2.35 keV at 1.33 MeV). The fully coded field of view (FCFOV) of the instrument 
is $16^\circ\times 16^\circ$ (Vedrenne et al. 2003). The imager 
IBIS has two detector layers: ISGRI and PICsIT. It has an angular resolution of 12 arc minutes and a 
FCFOV of  $9^\circ\times 9^\circ$ (Ubertini et al. 2003). The JEM-X monitor 
consists of two identical coded-aperture mask telescopes co-aligned with the other instruments, each 
with an energy  resolution of  $\Delta E/E = 0.47\times (E/1{\rm keV})^{-1/2}$ and an angular resolution 
of 3 arc minutes. Its FCFOV  is $4.8^\circ$  (Lund et al. 2003).

During the {\it INTEGRAL} Performance and  Verification Phase (hereafter PV, from  launch on October 17, 2002  to the  end of December 2002), an outburst of  
EXO 2030+375 was observed.  In the study presented here  we have \linebreak analysed  SPI, ISGRI and JEM-X data from revolutions  
18 to 23 (3--18 December 2002) around the Cygnus region, where this source is  
located. IBIS/ISGRI  was able to detect this source in \linebreak almost 285 pointing exposures made by the 
satellite.\linebreak However, only those pointings giving a detection level above $8\sigma$ have been taken into 
account ($\sim$ 170 pointings). SPI data almost cover  the duration of the outburst  which lasted approximately 
12 days ($\sim$ 450 pointings).  However, SPI data from revolution 18 was excluded  since
it was only a staring observation.  Due  to  the  pattern of observation exposures made by  {\it INTEGRAL}, \linebreak 
EXO 2030+375 was outside  the  JEM-X  field of view during the second part of the outburst, being detected in only 10 pointings. 

In addition, six more outbursts have been partially detected during 
{\it INTEGRAL} Galactic Plane Scans \linebreak(hereafter GPS) surveys. The period analysed includes revolutions 26 to 
269 (MJD 52630 -- MJD 53365). We have obtained positive detections with SPI in 16  revolutions,\linebreak with a 
total observing time of $\sim$ 770 ks. ISGRI has detected it within the fully coded field 
of view in 12 revolutions ($\sim$ 50 ks), and  JEM-X in 4 (9 ks) (see Tab. 1). 

{\it INTEGRAL} data reduction was carried out with ISDC's (Courvoisier et al. 2003)  Offline Scientific 
Analysis software, release 4.2.  A software description can be found in Goldwurm et al.  (2003),  
Diehl et al. (2003),  Westergaard et al. (2003). 

In 1996 July (MJD 50265--50275), 2002 June (MJD 52425--52446), and  2003
September (MJD 52894--52899), EXO 2030+375 was observed with the {\em RXTE} 
Proportional Counter Array (PCA) and the High Energy Timing Experiment (HEXTE)
\footnote{See http://heasarc.gsfc.nasa.gov/ for observation details}. For each
observation, we analysed PCA Standard 1 data (0.125 sec time resolution, no 
energy resolution) for light curves and Standard 2 data (16 sec time resolution,
129 channel energy resolution) for hardness ratio and spectral analysis using 
FTOOLs v5.3.1. Details of timing analysis of these data are given in Wilson et 
al. (2002, 2005).

\section{DATA ANALYSIS}

\subsection{Imaging}

Figure 1 shows the 5$\times$5 and hexagonal dithering patterns carried out by {\it INTEGRAL} during the PV 
phase in the Cygnus region, together with  the fully coded field of view of the three high-energy instruments. We 
observed that the 450 pointings can be grouped into a subset of 75 ($5\times 5 \times 3$) which are independent, due to 
repetitions of the observation pattern at the same location.

Table 1 shows the observations per instrument,  and  the mean flux in different energy ranges per 
revolution.  PV phase fluxes in the 20--40 keV energy range  obtained by ISGRI and SPI increase 
from MJD  52618 (revolution 19), peaking at around  MJD 52622 (revolution 20), and finally  
decreasing  to the end of the outburst at  MJD 52630 (revolution 23).
 
Using PV phase data for a total observing time of 306 ks  and  covering revolutions 18 to 22, \linebreak ISGRI 
located EXO 2030+375 at R.A.=308.09$^{\circ}$ and  DEC=37.65$^{\circ}$, with a 0.03$^{\circ}$ error 
radius.  SPI detected this source with a statistical significance of 35$\sigma$, averaging over  
revolutions 19 to 23 (730 ks), at  R.A.=308.12$\pm$0.07$^{\circ}$  and  DEC=37.47$\pm$0.05$^{\circ}$.  Only 
in 10 pointings of revolutions 19 and 20 (18 ks)  could JEM-X detect EXO 2030+375,\linebreak because it has the smallest 
fully coded field of view.  Its best location was at R.A.=308.06$^{\circ}$ and  DEC=37.64$^{\circ}$ 
with an error radius  of 0.03$^{\circ}$.

\begin{table}
  \caption{Journal of Observations}
  \label{JO}
  \begin{tabular}{lclll}
   \hline 
   \hline 
   \noalign{\smallskip}
    Instrument&Rev&Obs.Time&Mean MJD&Mean Flux \\
     		  &   & ks  & & mCrab\\
   \noalign{\smallskip} \hline \noalign{\smallskip}
   JEM-X          & 19 &  9 & 52618.73 	& 55$\pm$2\\
   (5-25 keV)     & 20 &  9 & 52621.72 	& 54$\pm$1\\
		  & 80 &  3.6 &52801.47	& 47$\pm$1\\
		  & 82 &  1.8 &52805.92 & 72$\pm$2\\
		  & 265&  1.8 &53353.52 & 88$\pm$2\\
		  & 266&  1.8 &53356.50	& 97$\pm$2 \\
   \noalign{\smallskip} \hline \noalign{\smallskip}
   IBIS/ISGRI & 18 & 19.8    & 52615.72  & 21.2$\pm$0.7\\
   (20-40 keV)& 19 & 104.4   & 52618.73  & 67.1$\pm$0.8\\
              & 20 & 113.4   & 52621.72  & 74.3$\pm$0.7\\ 
              & 21 & 63    & 52624.71  & 59$\pm$1\\
 	      & 22 & 5.4   & 52627.73  & 29$\pm$2\\
              & 52 & 4.4   & 52717.31   & 61$\pm$5\\
              & 67 & 2.2   & 52761.36   & 99$\pm$4\\
              & 80 & 18    & 52801.19   & 50$\pm$1\\
              & 82 &  2.2   & 52806.05   & 93$\pm$4\\
              & 142 & 4.4   & 52985.52   & 96$\pm$3\\
              & 145 & 4.4   & 52994.51   & 56$\pm$3\\
              & 162 & 2.2  & 53045.59   &  16$\pm$4\\
              & 189 & 2.2  & 53126.55   &  83$\pm$4\\
              & 193 & 2.2  & 53137.40   &  36$\pm$4\\
              & 265 & 2.2  & 53353.52   &  115$\pm$4\\
              & 266 & 2.2  & 53356.49   &  137$\pm$4\\
              & 269 & 2.2  & 53365.25   &  125$\pm$5\\
   \noalign{\smallskip} \hline \noalign{\smallskip}		
   SPI    	  & 19 & 187.2   & 52618.73  & 52$\pm$3\\  
   (20-40 keV)    & 20 & 178.2   & 52621.72  & 70$\pm$3\\
                  & 21 & 187     & 52624.71  & 45$\pm$4\\
	 	  & 22 & 206.8   & 52627.73  & 36$\pm$3\\
                  & 23 & 204.6   & 52630.79  & 6$\pm$2\\ 
 		  & 54 & 50      & 52722.89   & 32$\pm$14\\
		  & 59 & 46.2    & 52737.65   & 34$\pm$13\\
		  & 67 & 104.4   & 52761.36   & 78$\pm$12\\
		  & 79 & 17.6    & 52797.47   & 6$\pm$3\\ 
                  & 80 & 203     & 52801.19   & 52$\pm$2\\ 
                  & 82 & 28.6    & 52806.05  & 65$\pm$10\\
                  & 92 &  22     & 52835.92   & 26$\pm$15\\
                  & 142 & 24.2   & 52985.52  & 115$\pm$12\\
                  & 145 & 28.6   & 52994.51  & 22$\pm$12\\
                  & 189 & 41.8   & 53126.55  & 69$\pm$10\\
                  & 210 & 26.4   & 53188.42  & 36$\pm$13\\
                  & 218 & 35.2   & 53212.86  & 49$\pm$15\\
                  & 241 & 39.1   & 53281.68  & 37$\pm$11\\
                  & 253 & 40.3   & 53317.73  & 101$\pm$14\\ 
                  & 265 & 32     & 53353.52  & 115$\pm$17\\
                  & 266 & 28.1   & 53356.49  & 149$\pm$17\\
                  & 269 & 28.4   & 53365.25  & 118$\pm$15\\
   \noalign{\smallskip} \hline \noalign{\smallskip}
    \hline \noalign{\smallskip} 
   \end{tabular}
\end{table}

%-----------------------------------------------------------------------------------

%_____________________________________________________________
%                        figure with caption on the right side
%-------------------------------------------------------------
 \begin{figure}
  \centering
  \includegraphics[height=4cm,width=4.35cm]{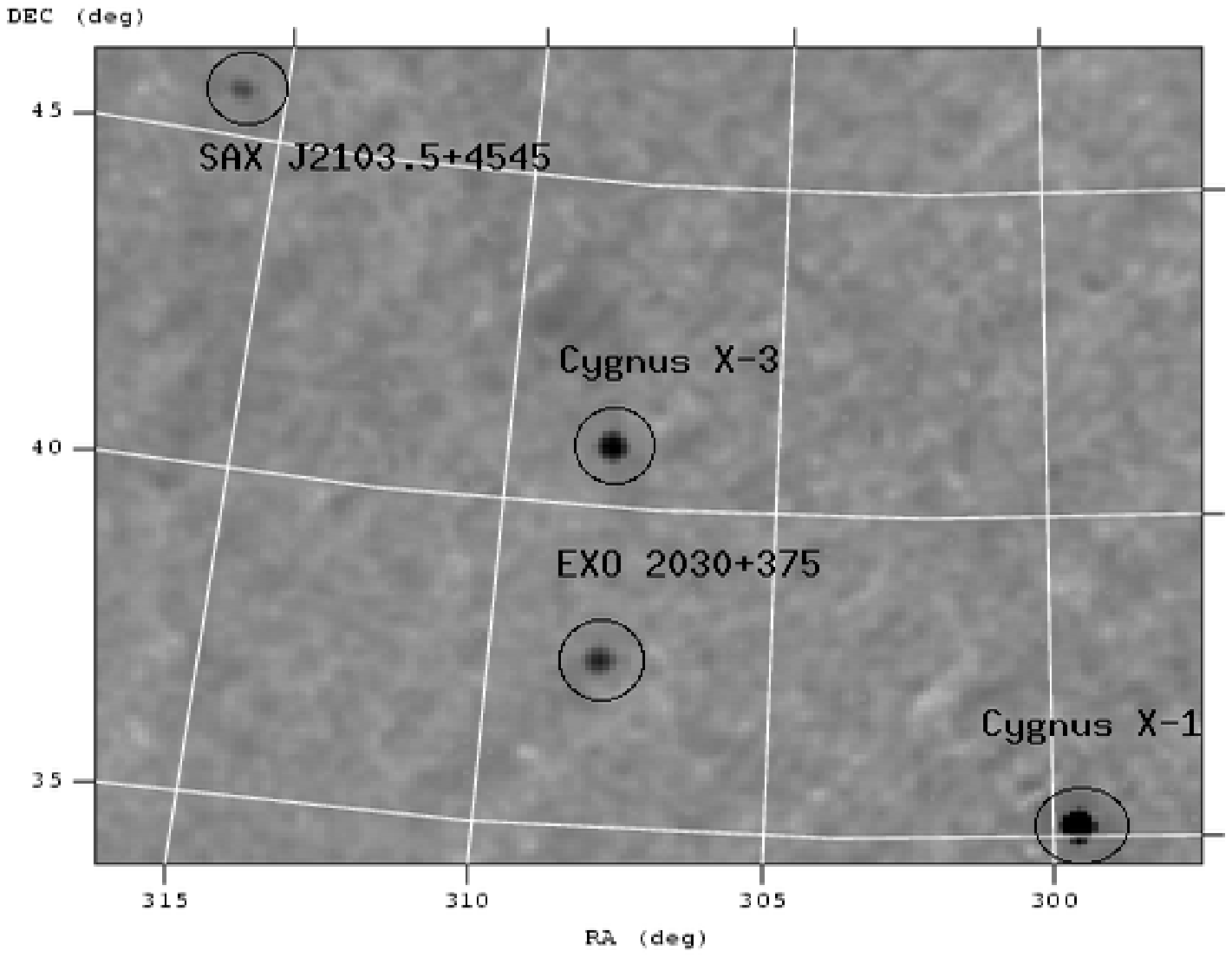}
  \includegraphics[height=4cm,width=4.35cm]{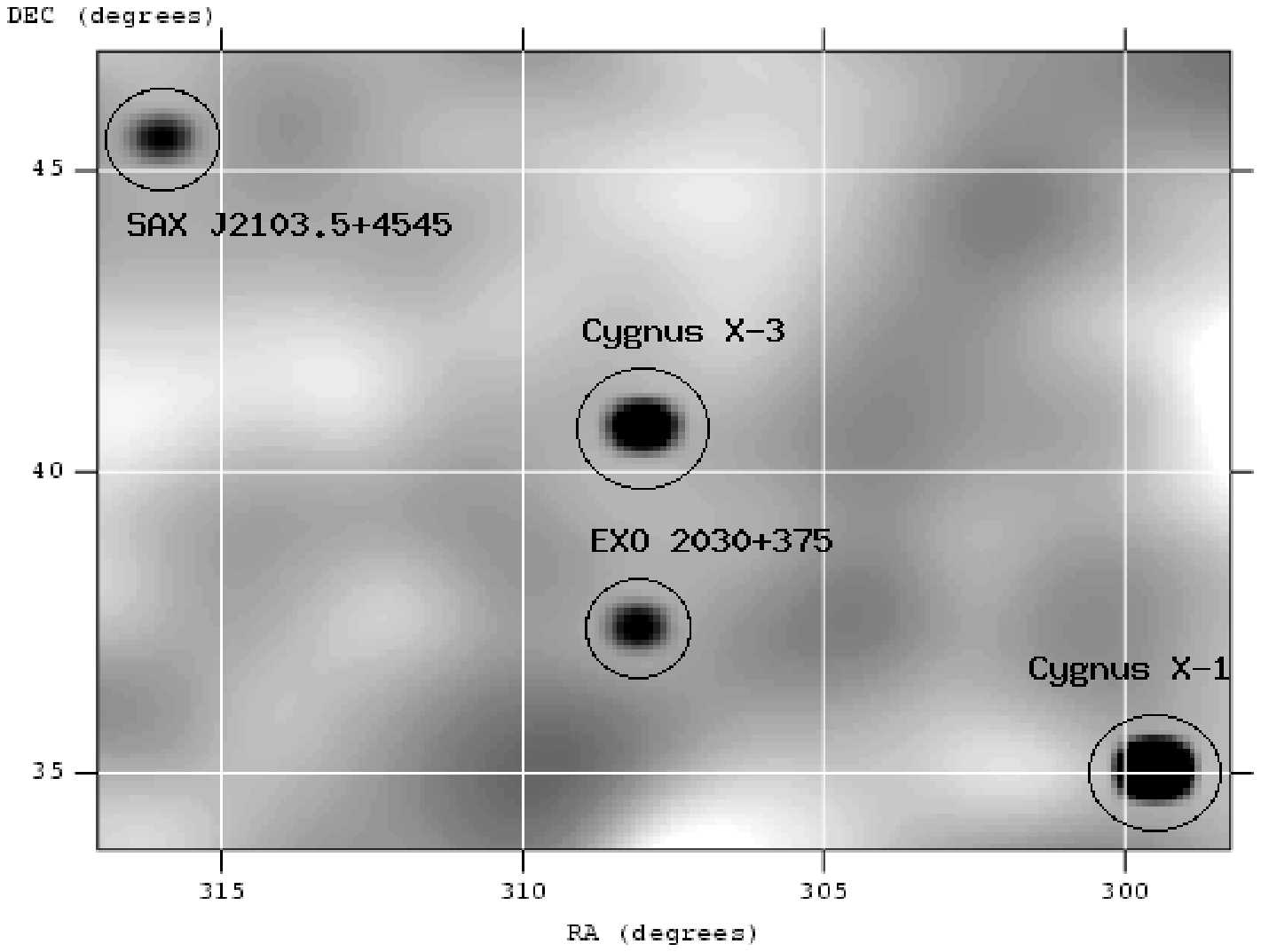}
    \caption{Left: a mosaic of the Cygnus region by ISGRI (revolution 20, 20--40\r\,keV).
             Right: an average image from revolutions 19 to 23 by SPI (20--40\rm\,keV).\label{imag}}
 \end{figure}

%----------------------------------------------

%--------------------------------------------------------------------------

 \begin{figure}
\includegraphics[height=6.8cm,width=8.7cm]{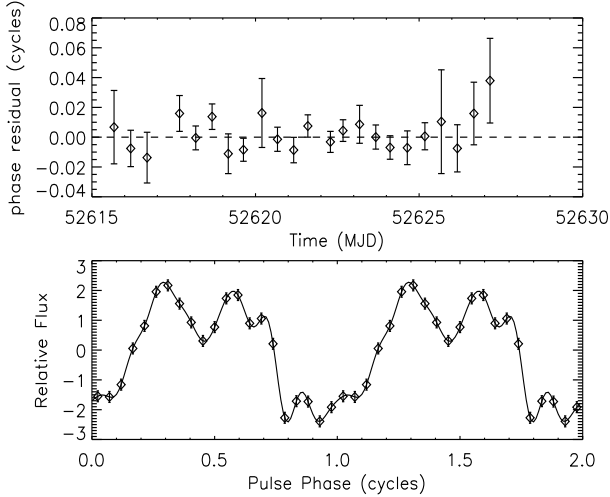}
\caption{Top: Pulse phase residuals for EXO 2030+375 ISGRI 20-60 keV data from 
revolutions 18-22 from a fit to a linear phase model. Bottom: Mean pulse profile
from revolutions 18-22 generated by combining profiles from individual science
windows using our linear pulse phase model.}
\label{fig:phases}
\end{figure}
 %--------------------------------------------------------------------------

\subsection{Timing}

X-ray pulsations have been detected by  JEM-X and IBIS/ISGRI. A standard 
epoch-folding analysis on JEM-X data gives a value of 41.601$\pm$0.005 s for the short-term 
variability. In addition, we have analysed ISGRI data using  new software based on previous 
experience in the timing analysis of X-ray pulsars with BATSE (\cite{Finger99}, \cite{Wilson02}). This 
software is able to generate a pulse phase for each science window and then fit them to get a 
pulse period. The best-fit period is  $41.691798 \pm 0.000016$ s (epoch MJD 52621.6913), 
using data spanning revolutions 18 to 22 (MJD 52614.419-52628.964), including barycenter (\cite{Walter03}) and 
orbit corrections (\cite{Wilson05}) . This result is in very good agreement with 
previous \linebreak results (Parmar  et al. 1989, Reynolds  et al. 1993,  Reig \& Coe 1998,  
Kuznetsov et al. 2004, Wilson et al. 2002, \cite{Wilson05} ).

For each science window, our software collected good events in the
partially and fully coded field-of-view of ISGRI in the 20-60 keV band using the
pixel information function (PIF). The good events were then epoch folded using 
initially a simple pulse phase model based on spin-frequency measurements from 
{\em RXTE}. To remove effects of binning, we then fit each folded profile with 
a Fourier series of harmonic coefficients. A template profile was then created 
from the average profile from the set of science windows. For each harmonic 
coefficient, we computed a reduced $\chi^2$.
\begin{equation}
\frac{\chi^2_h}{\rm d.o.f} =  \sum_{i=1}^N \frac{1}{(2N-2)} \left\{ \frac{(a_{ih} - \bar a_{h})^2}
 {\sigma_{a_{ih}}^2} + 
  \frac{(b_{ih} - \bar b_{h})^2}{\sigma_{b_{ih}}^2} \right\}
\end{equation}
where $a_{ih}$ and $b_{ih}$ are  the cosine and sine coefficients for science 
window $i$ and harmonic $h$, $\sigma_{a_{ih}}$ and $\sigma_{b_{ih}}$ are the
errors on those coefficients, $\bar a_h$ and $\bar b_h$ are the cosine\linebreak and sine
coefficients for the mean profile, and $N$ is the number of science windows.
To account for excess aperiodic noise from Cygnus X-1, since it cannot be
completely removed from the data, the errors on the harmonic coefficients for 
the individual science windows were multiplied by 
$(\chi^2_h/{\rm d.o.f})^{1/2}$. A similar technique was also applied to account
for excess aperiodic noise in BATSE data (\cite{Wilson02, Finger99}). To generate
phase offsets from the model, we then cross-correlated the individual profiles 
with the template profile.  The new phases (model + offset) were then fitted with a linear \linebreak or 
quadratic phase model, and the process was repeated,\linebreak creating new folded profiles, new harmonic 
coefficients, and new phase offsets. The pulse profiles were then combined over time using the 
phase model to improve statistics and allow the phase measurements to constrain \linebreak spin-up during the
outburst. Figure~\ref{fig:phases} shows 0.5 day phase residuals for our best fit 
model with a constant period,\linebreak indicating that no significant spin-up (or
spin-down) \linebreak was detected during the outburst. All of the individual pulse
profiles were combined using our phase model to produce the mean 20-60 keV
profile shown in the bottom panel of Figure~\ref{fig:phases}.

\begin{figure} 
  \centering 
  \includegraphics[height=8cm,width=8cm]{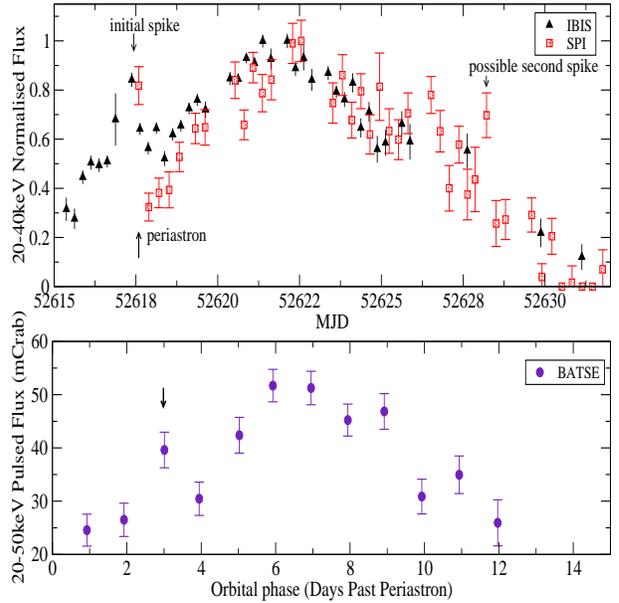} 
   \caption{ Top Panel: {\it INTEGRAL} (ISGRI--SPI, 20--40 keV) light curves of EXO 2030+375 
outburst seen in Dec. 2002. The unusual features are marked in both light curves.  Bottom 
Panel: BATSE 1-day average pulsed flux light curve (20-50 keV) from an outburst seen in 1993 
May. Fluxes were generated as described in Wilson et al. (2002). Only a marginal detection of the 
initial spike is observed. } 
   \label{lc} 
\end{figure}

%-------------------------------------------------

\begin{figure}
  \centering
  \includegraphics[height=7cm,width=8.8cm]{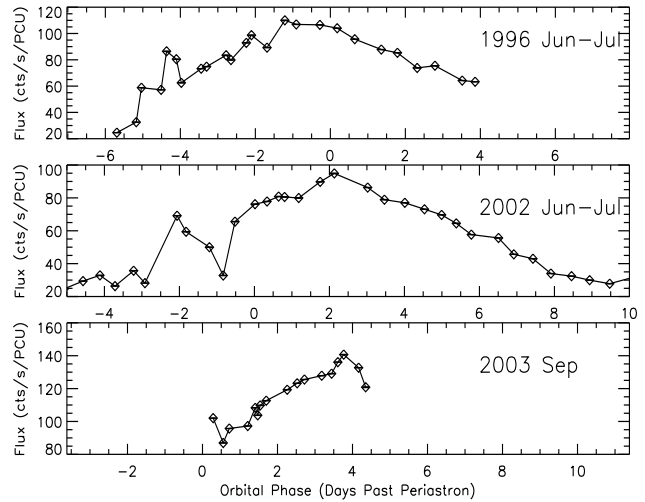}
   \label{lc3}
\caption[]{{\em RXTE} PCA light curves (2--60 keV) from 3 outbursts of EXO
2030+375. Evidence for an initial peak preceding the main outburst is present
in all 3 outbursts, although coverage for the third outburst is incomplete. No
secondary peak was seen.}
   \label{lc3}
\end{figure}

\begin{figure}
  \centering
  \includegraphics[height=6cm,width=9cm]{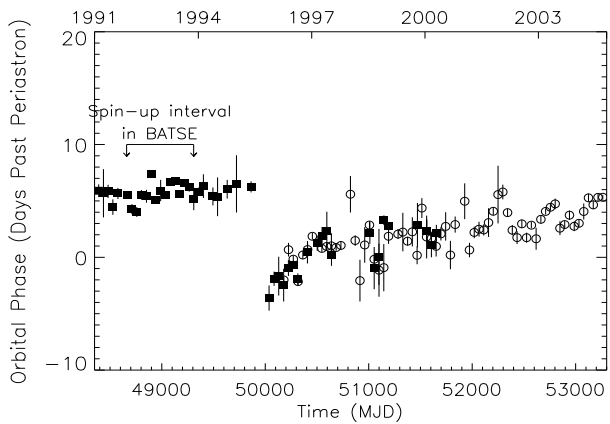}
\caption[]{Orbital phase of EXO 2030+375 main outburst 
peak versus time. Outburst peak times were measured with BATSE (filled squares)
and the {\it RXTE} ASM (open circles).}
\end{figure}

In the medium-term EXO 2030+375 shows type I \linebreak outbursts, i.e. increases of the 
X-ray flux during the \linebreak periastron passage of the neutron star.  During 
brighter EXO 2030+375 outbursts observed with BATSE,  Wilson et al. (2002) found a 
correlation between spin-up rate and pulse flux, suggesting that an 
accretion disk was likely to be present. The behaviour of this active period can be seen in Fig. 4, 
where a comparison  of the  (20--40\rm\,keV) ISGRI and SPI light curves  with 
that of an earlier outburst detected by BATSE (20--50 keV) is shown. The shape and  amplitude of variability 
in both cases are analogous, as well as the duration of the phenomenon (about 12 days). 
However, it is to be noted that in the same figure the {\it INTEGRAL} light curves show  a 4 day delay of  the maximum\linebreak of luminosity 
after the periastron passage of the neutron star, while the earlier outburst detected by BATSE peaked 6 days after periastron.

Furthermore, Fig. 4 shows  very clearly  an initial spike in both {\it INTEGRAL} light curves, and another possible spike after the outburst maximum. BATSE data shows marginal evidence for only an initial spike  about 3--4 days before the  maximum.   {\it RXTE}/PCA light curves from June-July 1996, 
June-July 2002, and September 2003 all show evidence for an initial peak preceding the maximum, 
but no evidence for a second spike (see Fig. 5). In these 3 outbursts, the 
initial spike, dip,  and maximum each appear to occur at approximately the same relative orbital phase, despite 
the fact that the orbital phase of the maximum shifts dramatically from about 1 day before periastron in 
June-July 1996 to about 2 days after periastron in  June-July 2002 and then to about 4 days after periastron 
as of September 2003 (See Fig. 6).

Galactic Plane Scans (GPS) performed by {\it INTEGRAL} allowed us to partially detect six more  
outbursts of this source. Figure 7 shows SPI, ISGRI and JEM-X light curves obtained with  GPS and PV phase data. For comparison {\it RXTE}/ASM data covering the same period are also plotted. In all light curves the seven outbursts seem to peak at the same time, following the same trend as previous type-I outbursts: the same duration and shape, and all separated by about 46 days, which corresponds to the orbital period of the system. However, the amplitude of variability varies from one to another. In principle, this can be due to the outbursts  not being completely sampled intime during GPS, since the purpose of these observations is to perform saw-tooth-path  
scans of the Galactic Plane at weekly intervals, where  individual exposures are separated by $6^\circ$ along the scan path. On the other hand, due to its  detector design, data from  SPI are highly background dominated. For SPI to achieve a signal to noise ratio of  $\sim$10 for a source as bright as  EXO 2030+375 an observation of 300 ks would be necessary. Therefore the GPS pointing pattern is not ideally suited to make a detailed study of point sources.

%---------------------------------------------

%-------------------------------------------------------------- 
 
\begin{figure} 
  \centering 
  \includegraphics[height=5.7cm,width=8.7cm]{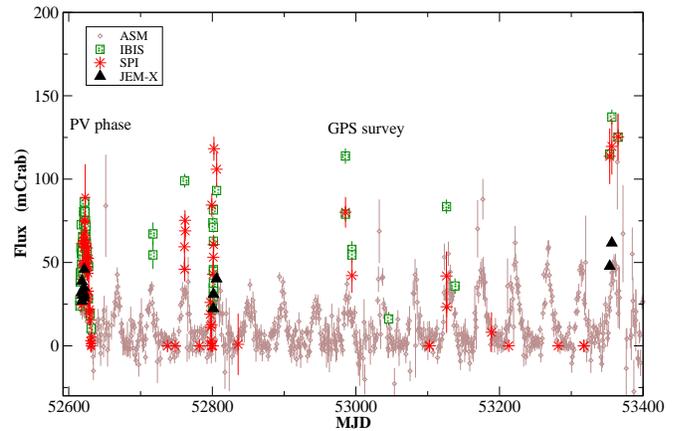} 
   \caption{ {\it INTEGRAL} (5--25\rm\,keV JEM—X, 20—-40\rm\,keV ISGRI and SPI) light curves from different outbursts, and simultaneous data from {\it RXTE}/ASM (1.5--12\rm\,keV).}
   \label{lc} 
\end{figure} 
 
%--------------------------------------------------------------- 
 
 \subsection{Hardness Ratios}

In order to  study  the spectral variability of EXO 2030+375  we have carried out a hardness ratio ({\it HR}) analysis, since 
the quality  the data prevent us from performing  a detailed analysis.  Table 2 lists the energy bands used. The HR is defined as:
 \begin{equation}
 HR =  \frac{H-S}{H+S} 
\end{equation}
with H and S being the hard and soft band fluxes, respectively.

Figure 8 (top) shows  {\it RXTE}/ASM, JEM--X, ISGRI and SPI  HRs during the PV phase outburst. All data 
are in Crab units. In spite of the poor statistics, at high \linebreak energies  there seems to exist a 
softening of the spectrum (ISGRI and SPI data), while at low energies a hardening \linebreak is seen ({\it RXTE}/ASM data). 
It should be noted that for JEM--X only the first half of the outburst is present.

Figure 8 (bottom left) shows  JEM--X  (5--15\rm\,keV) and (15--25\rm\,keV) light curves. We cannot clearly see an 
increase of flux at higher energies. On the other hand, (20--40\rm\,keV) and 
(40--60\rm\,keV) SPI light curves (bottom right) show a decrease of the flux in the higher energy band.

Hardness ratios were also generated from {\em RXTE} PCA data using three energy
bands: 2-5, 5-15, and 15-30 keV. Here the hardness ratios were defined as $HR=H/S$. In all 
three outbursts observed with the {\em RXTE} PCA, the flux in the 5-15 keV band was larger than that in the other
two bands. This differs from where the spectrum peaked in the {\em INTEGRAL} 
data, but is likely related to differences in the spectral response of the 
instruments.

During the main part of all three outbursts, the hardness ratios were roughly 
constant, as was also observed with {\em INTEGRAL}. However, the 5-15 keV/2-5 keV 
ratio, shown in Figure~\ref{fig:rxtehr}, exhibited intriguing behavior in the 
dip preceding the main outburst. In the 1996 outburst, the dip is harder than
the main outburst, while in both the 2002 and 2003 outbursts, the dip is softer than
the main\linebreak outburst. Also, this hardness ratio was correlated with\linebreak intensity in 
the 2002 outburst, when our observations \linebreak covered the largest range of orbital
phases and intensities.

%------------------------------------------ 
\begin{table} 
\centering  
  \caption{Hardness ratios energy bands per instrument.}
  \begin{tabular}{lll} 
   \noalign{\smallskip} 
    \hline \hline \noalign{\smallskip}
    Instrument   &  Hard enery band &    Soft energy band \\
    \hline \noalign{\smallskip}
     {\it RXTE}--ASM   &  5--12 keV    & 3--5 keV\\
    \hline \noalign{\smallskip} 
     JEM--X      &  15--25 keV   & 5--15\rm\,keV\\
    \hline \noalign{\smallskip} 
     ISGRI       &  40--60 keV    & 20-- 40\rm\,keV\\
    \hline \noalign{\smallskip} 
     SPI         &  40--60 keV    & 20-- 40\rm\,keV\\ 
    \hline \hline \noalign{\smallskip} 
   \end{tabular} 
   \label{broad} 
% \end{center} 
\end{table}

\begin{figure} 
  \centering 
  \includegraphics[height=7.5cm,width=8.3cm]{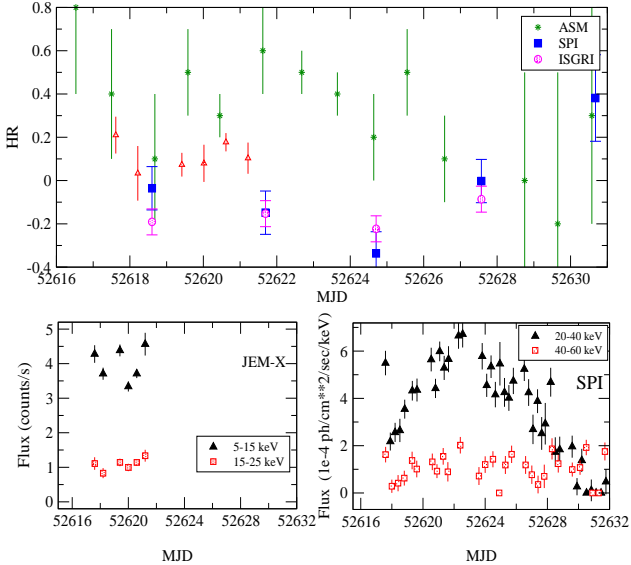} 
   \caption{Top: Hardness Ratios from {\it RXTE}/ASM, JEM-X, ISGRI and SPI during PV phase outburst. 
     Bottom left: (5--15 keV) and (15--25 keV) JEM-X  light curves. Bottom right: (20--40 keV) and 
    (40--60 keV) SPI light curves. At high energies a softening  of the spectrum is seen.}
   \label{lc} 
\end{figure} 

\begin{figure}
\includegraphics[height=6cm,width=8.5cm]{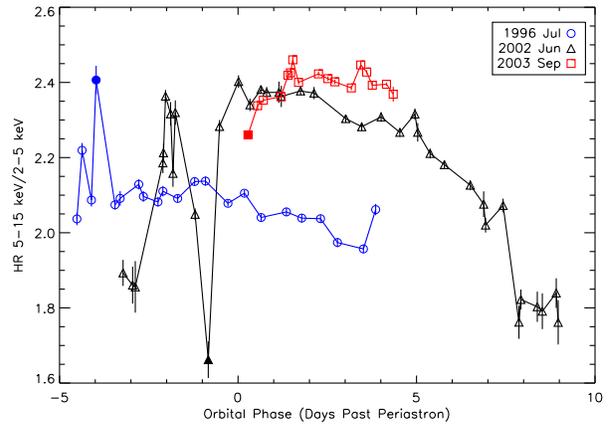}
\caption{{\em RXTE}/PCA 5-15 keV/2-5 keV hardness ratio vs. orbital phase from
three outbursts of EXO 2030+375. Filled symbols denote the hardness ratio for
the dip preceding the main outburst.\label{fig:rxtehr}}
\end{figure}

\subsection{Spectral analysis}

In this section we have characterised the average spectral shape of EXO 2030+375 during the December 2002 \linebreak outburst.
Data from the three high-energy instruments have been combined to achieve a 5--300 keV 
broad-band average spectrum (Fig. 10). It is the first time that such a broad-band spectrum of this source 
is presented. For JEM--X we have obtained an average spectrum in the \linebreak 5—-25 keV energy range  by using 
five Science Windows of revolution 19 (JEM--X 2 data). Only IBIS/ISGRI spectral \linebreak data within the fully coded field of view were used  for \linebreak extracting the 20--100 keV average spectrum. This \linebreak
includes 35 spectra from revolutions 19 and 20.  SPI has the \linebreak largest field of view, hence EXO 2030+375 was detected throughout almost the entire duration of the outburst.  In order to improve the signal to noise, data from revolutions 19 to 23 were used for attaining an average spectrum in the 20--300 keV energy range.  
 
We found that the broad-band spectrum can be \linebreak adequately described by the sum of a disk black body  
with either a power law model  or a Comptonized component\linebreak (COMPTT; Titarchuk et al. 1994, 1995) . The soft 
component is interpreted as coming from the accretion disk, the neutron star surface, or an optically thick 
boundary layer (Barret et al., 2000 and references therein). The harder one is often modeled by simple power 
laws, however Comptonization models provide more physical fits. This process is speculated to take place in a 
scattering corona located somewhere in the system: around the neutron star (e.g., optically thin boundary layer, 
spherical corona) or above the disk.
 
Table 3 summarises the spectral parameters that have been obtained fitting both models. The soft component has a
black body temperature of kT$_{BB}\sim$8 keV,  which  does not substantially change after applying a power law or 
a COMPTT model. In addition,  we can see that an unbroken power law fits well the hard X--ray component with a 
$\Gamma$=2.04$\pm$0.11 keV. Moreover, we found compatible \linebreak fits when a COMPTT model is used  with different\linebreak electron 
temperatures between kT$_e$=30 keV and \linebreak kT$_e$=60 keV, although it was not possible to constrain \linebreak this 
parameter,  hence  it was fixed in all the fits. We \linebreak noticed how the optical depth decreases as the temperature
increases, while the rest of the parameters are almost constant with temperature.

We have applied both spherical and disk geometries yielding to  different values for the optical depth of 
the electron cloud ($\tau_{sphe}$/ $\tau_{disk}$ $\sim$2). However, the temperature \linebreak of the seed photons (kT$_W$),  
assuming a Wien-type \linebreak distribution, as well as the black body temperature\linebreak remained constant.

In any case the addition of an absorption component improved the fits  (a value of N$_H=(2.6\pm 0.3)\times 10^{22}$ cm$^{-2}$ was 
used; Reig \& Coe 1999).  Neither an iron line was detected nor any cyclotron line feature around 36 keV as Reig et al. (1999) suggested 
could be observed.  No cut-off\linebreak at 200--300 keV, as seen in black-hole systems (BHs) like Cygnus X—-1 (Sch$\ddot{o}$nfelder 2001), seems 
to be present\linebreak although  the data are not conclusive. 

%------------------------------------ 
 
\begin{figure} 
  \centering 
  \includegraphics[height=6cm,width=8.9cm]{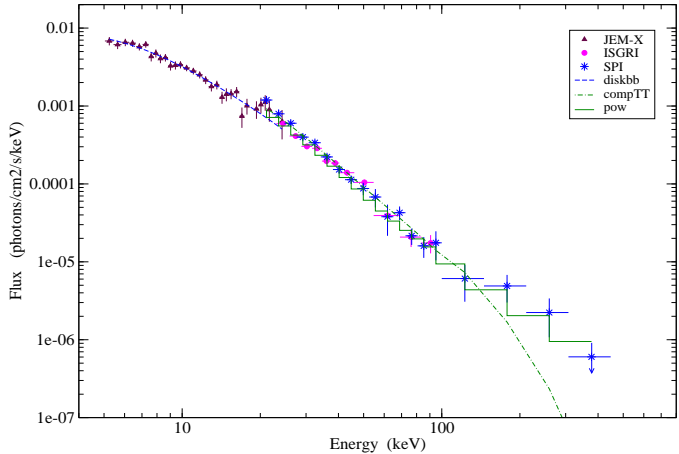} 
  \caption{Broad-band spectrum obtained combining JEM--X-ISGRI-SPI mean spectrum of 
EXO 2030+375. Two models have been fitted to the data: a disk black body plus either a power law (stepped line) or a comptonized component of kT$_e$=30 keV (dashed line).} 
  \label{index} 
\end{figure} 
 
%---------------------------- 

%------------------------------------------ 
\begin{table} 
\centering  
  \caption{Broad-band spectral parameters (5--300 keV) with errors estimated in the 90\% confidence range . For 
the comptonized component a spherical geometry was selected.} 
  \begin{tabular}{lll} 
   \noalign{\smallskip} 
    \hline \hline \noalign{\smallskip}
    Model& Parameters            & Intercalibration \\
         &                       & factors\\
         &                       &(ISGRI/JEM-X) \\
         &                       &(SPI/JEM-X)\\
\hline \noalign{\smallskip}

                   & kT$_{BB}$= $8.2{+1\atop -0.9}$ keV   & 1.0${+0.2\atop -0.1}$\\ 
    diskbb         & $\alpha$= 2.04$\pm$0.11  			& 1.1$\pm$0.2 \\
     + pow         & $\chi^2_{red}$/dof= 1.07/72    		&\\
                   & Flux$^*$= 1.7 $\times10^{-9}$ 		&\\
    \hline \noalign{\smallskip} 
 
                   & kT$_{BB}$= 8$\pm$2 keV   & $1.2{+0.3\atop -0.2}$  \\ 
    diskbb         & kT$_W$= $1.5{+0.3\atop -0.4}$  keV   & $1.3{+0.3\atop -0.2}$ \\
    +compTT        & kT$_e$= 60 keV  (fixed)   	 &\\
                   & $\tau$= 1.3${+0.4\atop -0.3}$  &\\
                   & $\chi^2_{red}$/dof= 1.005/70      &\\
                   & Flux$^*$= 1.5 $\times10^{-9}$    &\\
 \hline \noalign{\smallskip} 
 
                   & kT$_{BB}$=7$\pm$2 keV   & $1.18{+0.3\atop -0.18}$  \\ 
    diskbb         & kT$_W$= 1.5${+0.3\atop -0.4}$  keV   & $1.31{+0.18\atop -0.2}$ \\
    +compTT        & kT$_e$= 30 keV  (fixed)   	 &\\
                   & $\tau$= 2.6${+0.6\atop -0.5}$  &\\
                   & $\chi^2_{red}$/dof= 1.03/70      &\\
                   & Flux$^*$= 1.4$\times10^{-9}$    &\\
    \hline \hline \noalign{\smallskip} 
    $^*$(erg cm$^{-2}$s$^{-1}$) & &\\
   \end{tabular} 
   \label{broad} 
% \end{center} 
\end{table}

 \section{DISCUSSION}

We have carried out {\it INTEGRAL} imaging, timing and spectral analysis of the Be/X-ray binary EXO 2030+375 
December 2002 outburst. In addition, we have reported six more outbursts  detected during 
{\it INTEGRAL} GPS\linebreak surveys. 

Temporal analysis performed with JEM—X and IBIS/ISGRI observations showed X—-ray pulsations. Previous results using 
{\it INTEGRAL} data by Kuznetsov et al. (2004) gave a value of 41.6897$\pm$0.0001 s. Our  best fit period obtained 
is $41.691798 \pm 0.000016$ s,  in \linebreak excellent agreement with previous results (Parmar  et al. 1989, Reynolds  et al. 
1993,  Reig \& Coe 1998,  Wilson et al. 2002, \cite{Wilson05}). 
The global trend in the pulsar spin frequency was spin-down from 1994 through 2002. Recently, based on the 2002 and 
2003 outburst observations with {\it RXTE} PCA, Wilson et al. (2005) observed a change from spin-down to spin-up. 
They suggested that the pattern of constant spin, followed by spin-up,\linebreak followed by spin-down, observed with 
BATSE, is repeating \linebreak with an approximately  $\sim 11$ yr cycle. Nevertheless, no\linebreak significant spin-up (or spin-down) was 
detected during the {\it INTEGRAL} PV phase outburst.

The X-ray behaviour of EXO 2030+375 in the medium term, is characterised by a regular increase of the\linebreak X-ray 
flux modulated with the orbital period (type I \linebreak outbursts).  The average X-ray luminosity was\linebreak 
9.7$\times10^{36}$ erg s$^{-1}$,  for an assumed distance of 7.1 kpc \linebreak(Wilson et al. 2002). 
This luminosity value  is of the  same order as previous type--I outbursts of this source. 
We have estimated the L$_{1—-20}$/L$_{20-200}$ ratio for EXO 2030+375, obtaining a  value  $\sim$1.2. This 
allows us to locate this source in the so-called X--ray burster box, whereas all black holes
are found outside it (Barret et al. 2000).

During the  PV phase outburst a four-day delay \linebreak between the maximum of luminosity and the periastron passage 
of the neutron star was found.  Previous results showed that EXO 2030+375's outbursts abruptly shifted from 
peaking about 6 days after periastron to peaking before periastron and then gradually shifting to after 
periastron (see Fig. 6), depending on a density perturbation precessing in a prograde direction around \linebreak
the  Be disk (Wilson et al.  2002). Recent {\it RXTE} ASM data indicate that the main peak is currently about 5 
days after periastron.

The second  feature found in the EXO 2030+375 {\it INTEGRAL}/SPI light curve is not detected by {\it RXTE}, and 
there are not enough ISGRI data covering this period for comparison.  When interpreting  SPI timing analysis 
it is important to take into account the  possible influence of a very strong source like Cygnus X--1 in the field of view 
of this coded mask instrument. To resolve all sources each should have a clear coding pattern from the mask, and if 
this is incomplete for any of the sources, especially for a weak one,  the mask pattern may correlate with parts of the 
mask pattern of neigbouring sources introducing a \linebreak systematic error in flux values. This phenomenon of an incomplete detector 
pattern due to a  source \linebreak outside the FCFOV can also be seen in the two dips at the end of the Cygnus X--1 light curve. The observation 
detector pattern of the second EXO 2030+375 feature is not completely inside the FCFOV of the instrument, i.e. this feature could have a  
systematic error due to cross correlation with Cygnus X--1.  In Fig. 11  we can see  four large flares in the Cygnus X--1 
light curve. In principle, if such a kind of correlation is occurring  we would expect \linebreak four unusual features in 
EXO 2030+375 light curve. However, only the first and last of these correlate to the EXO 2030+375 features, but 
there is no correlation for the second and third. Therefore, we cannot rule out the possibility that the second  
peculiarity in the SPI data of EXO 2030+375 is real.

Analysis of {\it INTEGRAL} and {\it RXTE}  data did not show clear spectral differences between the initial and 
maximum  peaks.  If the emission characteristics during the initial peak  and the main outburst are  different this would 
indicate two different accretion mechanisms. Interestingly, the RXTE data showed evidence for a spectral difference in the 
dip between the peaks. However limited statistics did not allow a  detailed spectral analysis.

2S 1845-024, a 94-second pulsar in an eccentric (e=0.88) 242-day orbit, 
exhibits the same recurrent \linebreak structure consisting of an initial peak preceding 
the\linebreak maximum  and followed by another weaker peak (Finger et al. 1999). In this 
source, the spikes are interpreted as due to the passage of the pulsar through 
the Be star's\linebreak circumstellar envelope in an orbit  inclined with respect to the 
equatorial plane of the Be star, while the main peak was interpreted as
due to disk accretion.

\begin{figure} 
  \centering 
  \includegraphics[height=6cm,width=8.9cm]{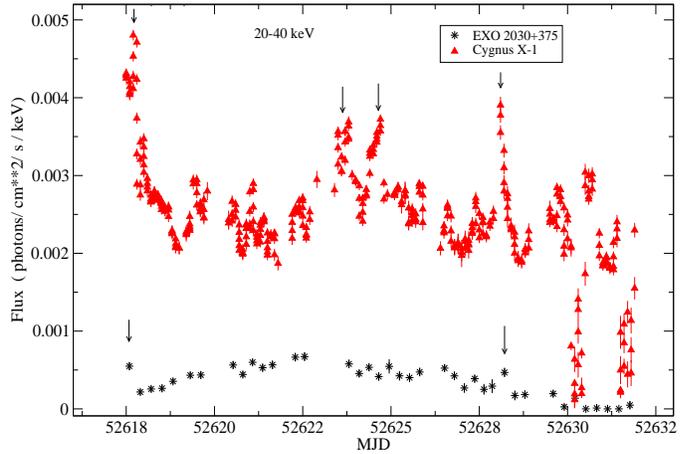} 
  \caption{Cygnus X--1 and EXO 2030+375 light curves from {\it INTEGRAL} PV phase data. The four maxima in the 
    Cygnus X--1 light curve and the two EXO 2030+375 spikes are marked with arrows.} 
  \label{index} 
\end{figure}

Reig \& Coe (1999) studied the continuum  spectral \linebreak shape of EXO 2030+375 in the 
2--20 keV energy range,  representing it  with an exponential cut-off
($\Gamma=1.00\pm 0.06$, E$_{cut}\sim 8.85\pm $0.33\rm\,keV), plus absorption by cold material 
(N$_H=(2.6\pm 0.3)\times 10^{22}$ cm$^{-2}$) and an iron emission line at  $\sim$ 6.5 keV.  
In the Gamma-ray band data are scarce. {\it CGRO}/BATSE observations in the 20--150 keV energy range 
did not find significant variations\linebreak of the spectral parameters over a 134-day period, with  a 
thermal bremsstrahlung as the best-fitting model \linebreak ($kT=20.2\pm 0.3\rm\,keV$, \cite{Stollber99}). Recent  
{\it INTEGRAL} observations in the 5--25 keV range by Mart\'{\i}nez N\'u\~nez et al. (2003) have reported  
as the best fit an exponential cut-off ($\Gamma=0.52\pm 0.12$, E$_{cut}\sim 7.8$\rm\,keV),\linebreak  althought an 
iron emission line was not detected. Kuznetsov et al. (2003) fitted the 20--100 keV spectrum \linebreak  using  a power 
law plus an exponential cut-off\linebreak  ( $\Gamma=1.5\pm 0.2$, E$_{cut}\sim 37$\rm\,keV). Furthermore, in the\linebreak  20--150 keV 
energy range Bouchet et al. (2003) found that a single power law ( $\Gamma=2.72\pm 0.02$) or a bremsstrahlung 
($kT=27.2\pm 0.1\rm\,keV$) were compatible fits. Preliminary 10--200 keV broad-band spectrum by Camero 
et al. (2004) showed  that either an unbroken power law \linebreak  ( $\Gamma=2.54\pm 0.14$) or a bremsstrahlung 
($kT=34\pm 0.4\rm\,keV$) were again good fits.
In the present  work the hard component of the  5--300 keV broad-band spectrum,  described  by  a single power 
law model with a $\Gamma$=2.04$\pm$0.11 keV, confirms {\it INTEGRAL} preliminary results. As we said in the previous 
section and in order to provide a physical fit for EXO 2030+375 we have used in addition a Comptonization 
model (COMPTT). Due to the fact that this is the first attempt made for this source we can not compare our results 
with previous studies. However, we will compare them with other analysis performed in another neutron stars and 
black-hole systems. These studies, however, are concentrated on NS in low-mass systems with lower magnetic fields than 
accreting X-ray binaries. Such detailed studies using physical models for broad-band spectra  have not been done 
frequently  for accreting pulsars.

%\textbf{ %

On average, BH spectra are harder than NS\linebreak  spectra and this has been tentatively explained 
by the additional cooling provided by the NS surface, which may act as a thermostat capable of limiting the maximum 
kT$_e$ achievable in these systems (Kluzniak 1993; Sunyaev \& Titarchuk 1989). Using the COMPTT 
model for fitting BH spectra showed that electron temperatures greater than 50 keV seem to be common, while NS 
temperatures\linebreak  are usually smaller than that value (Barret  2001). For EXO 2030+375 we have found that 
electron temperatures between  30 keV and 60 keV  are compatible in describing\linebreak  the broad-band spectrum hard 
X—-ray component. The 60 keV temperature value  seems to be inconsistent with the  general behaviour of NSs. However, that  
criterion is still under debate since a temperature of about 50 keV,\linebreak  derived from the  fitting of the hard X--ray spectra with a 
Comptonization model,  was found by Torrej\'{o}n et al. (2004) for the High Mass X--ray Binary 4U2206+54, as well as some 
speculative BH candidates for which kT$_e$   is below 50 keV  (e. g., IGR J17464-3213 and  GRS 1758-258;\linebreak  kT$_e\sim$18 keV 
and $\sim$33 keV, respectively; Capitaino et al. 2004, Mandrou et al. 1994). In any case, this is \linebreak applied when an energy 
cutoff is observed. But such\linebreak  cutoffs are not always present in NS hard X--ray spectra. EXO 2030+375 does not show 
a cuttoff ( $\Gamma=2.04$). The same behaviour is found in some weekly magnetised NS like 1E 1724-3045 and Aql X-1, 
revealing a nonattenuated hard power law of $\Gamma=1.8$ up to 200 keV (Barret et al. 1991) and $\Gamma$ in the range 
2.1--2.6 in the 20-100 keV \linebreak energy range (Harmon et al. 1996), respectively. Therefore, the indication of a steep 
hard X--ray (E$\gae$30 keV) spectrum with $\Gamma\gae$ 2.5  cannot alone be used to claim that such a spectrum  comes from a 
NS. Similarly, a hard X--ray power-law spectrum with $\Gamma$$\lae$2.5 is not unique to BHs 
(e.g., 4U 0614+091: $\Gamma$=2.3; Piraino et al. 1999). Barret et al. (2000) suggested that there might be two 
classes of NSs. Members of the first class would display hard X--ray spectra with energy cutoffs, which would result \linebreak 
from thermal Comptonization. Members of the second class, where EXO 2030+375 could be included,  would have 
nonattenuated power laws which could be produced by nonthermal Comptonization, similar to the ones observed 
in the soft state of BHs.

%}

\section{CONCLUSIONS}

Our main results can be summarized as follows:\linebreak 
$\bullet$  we have obtained the first  5--300 keV broad-band \linebreak spectrum using {\it INTEGRAL} observations of the December 2002 
EXO 2030+375 outburst. A nonattenuated power law ($\Gamma=2.04\pm0.11$) or a COMPTT model (kT$_e$=30 keV) \linebreak can describe 
the hard X--ray component, while a disk black body can represent the soft component (kT$_{BB}\sim$8 keV).\linebreak  It should be  noted 
that this is the first attempt to apply a physical fit to EXO 2030+375;\linebreak $\bullet$ we have reduced pulse period uncertainty
determinations to $\sim$ 20\% of previous INTEGRAL measurements; \linebreak $\bullet$ in the medium-term analysis,  
two unusual features in the light curve were found, an initial peak before the main outburst and another  weaker one after the maximum. The physical mechanisms producing these features
are unknown. {\it RXTE} observations confirm only the existence of the initial spike.  Further observations would be needed to find out which physical mechanisms  modulate the X--ray flux. 
\linebreak
\linebreak
\linebreak
\linebreak
Acknowledgements. We thank Peter Kretschmar for very useful comments that helped to improve this paper and Julien 
Malzac for providing useful information related to the NS and BH broad-band spectral behaviour. We appreciate the 
interesting discussions with  Paco Bontempi and  his support. We thank Mark Finger for considerable help with 
the software used for ISGRI timing analysis. We also appreciate The Exploration of the Universe Division (EUD)at 
NASA's Goddard Space Flight Center for the opportunity to develop the present work. This research is supported by 
the Spanish Ministerio de Educaci\'{o}n y Ciencia (former Ministerio de Ciencia y Tecnolog\'{i}a) through 
grant-no ESP2002-04124-C03-02.

\end{document}